\documentclass[11pt,twoside]{article}

\usepackage{asp2006}
\usepackage{epsf}
\usepackage{lscape}

\begin{document}

\title{What hydrodynamical simulations tell us about the radial properties of
the stellar populations in Ellipticals}   %%% Fill in title

\author{Antonio Pipino,\altaffilmark{1,}\altaffilmark{2} 
Annibale D'Ercole,\altaffilmark{3} 
and Francesca Matteucci\altaffilmark{2} }  
%\altaffiltext{1}{Astrophysics, University of Oxford, Denys Wilkinson Building, Keble Road, Oxford OX1 3RH, U.K.}
%\altaffiltext{2}{Dipartimento di Astronomia, Universita di Trieste, Via G.B. Tiepolo 11, 34100 Trieste, Italy}
%\altaffiltext{3}{INAF-Osservatorio Astronomico di Bologna, via Ranzani, 1 - 40127 Bologna - Italy} 
\affil{$^{1}$Astrophysics, University of Oxford, Denys Wilkinson Building, Keble Road, Oxford OX1 3RH, U.K.}
\affil{$^{2}$Dipartimento di Astronomia, Universita di Trieste, Via G.B. Tiepolo 11, 34100 Trieste, Italy}
\affil{$^{3}$INAF-Osservatorio Astronomico di Bologna, Via Ranzani 1, 40127 Bologna, Italy}

\begin{abstract} %%% Abstract to run on from here.
Elliptical galaxies probably host the most metal rich stellar
populations in the Universe. The processes leading to both the formation and
the evolution of such stars are discussed by means of a new gas dynamical
model which implements detailed chemical evolution prescriptions.
Moreover, the radial variations in the metallicity distribution of these
stars are investigated by means of G-dwarf-like diagrams.

By comparing model predictions with observations, we derive a picture of
galaxy formation in which the higher is the mass of the galaxy, the shorter
are the infall and the star formation timescales.
The galaxies seem to have formed outside-in, namely the most external
regions accrete gas, form stars and develop a galactic wind very quickly (a
few Myr) compared to the central core, where the star formation can last up
to 1 Gyr.

We show for the first time a model able in reproducing the
mass-metallicity and the color-magnitude relations as well as
the radial metallicity gradient, and, at the same time, the observed
either positive or negative slopes in the [$\alpha$/Fe] abundace ratio gradient
in stars.
\end{abstract}

\section{The model}   %%% Top level section head (remove "%" symbol)

We present preliminary results from a new class of hydrodynamical models for the formation
of single elliptical galaxies (Pipino, D'Ercole, \& Matteucci, 2007) in which we implement detailed
prescriptions for the chemical evolution of H, He, O and Fe.  Our aims
are: i) to test and improve our previous predictions of an outside-in
formation for the majority of ellipticals in the context of the
supernovae-driven wind scenario, by means of a careful study of gas
inflows/outflows; ii) to explain the observed slopes, either positive
or negative, in the radial gradient of the mean stellar [$\alpha$/Fe],
and their apparent lack of any correlation with all the other
observables (Melhert et al. 2003; Annibali et al. 2006; 
Sanchez-Blazquez et al. 2007).

We adopted a one-dimensional hydrodynamical model which follows the
time evolution of the density of mass, momentum and
internal energy of a galaxy, under the assumption of spherical
symmetry.  In order to solve the equation of hydrodynamics with source
term we made use of an improved version of the Bedogni \& D'Ercole (1986) Eulerian,
second-order, upwind integration scheme. Our main novelty, in fact, is that
we follow the chemical evolution of several elements, namely H, He,
O and Fe. This set of elements is good enough to characterize our
simulated elliptical  galaxy from the chemical evolution point of view.  In
fact, as shown by the time-delay model (Matteucci \& Greggio 1986), 
the [$\alpha$/Fe] ratio is a powerful estimator of
the duration of the star formation.

\section{Results}

All the models run undergo an outside-in formation as suggested by Pipino, Matteucci, \& Chiappini (2006),
in the sense that star formation stops earlier in the outermost than
in the innermost regions, owing to the onset of a galactic wind.
We find that the predicted variety of the gradients in the [$\alpha$/Fe]  
ratio can be explained by physical processes, generally not taken into 
account in simple chemical evolution models, such as metal--enhanced 
radial flows coupled with different 
initial conditions.

We find [Fe/H] gradient slope in the range -0.5 -- -0.2 dex per decade
in radius and -0.3 dex per decade in radius for [Z/H], in
agreement with the observations (e.g. Annibali et al. 2007).  

Once transformed into
predictions on the line-strenght indices, these gradients in the
abundaces lead to $d \rm Mg_2/log (R_{eff,*}/R_{core,*})\sim -0.06$ mag
per decade in radius, again in agreement with the typical mean values
measured for ellipticals by several authors and confirming the
Pipino et al. (2006) best model predictions.  The remarkable exception of some model
with a steeper gradient seems to go in the direction of a few massive
objects in the Ogando et al. (2005) sample.

By analysing typical massive ellipticals, we find that all the models
that show values for their chemical properties, including the
[$\rm <Fe/H>_V$] and the total metallicity gradients, within the observed
ranges, show a variety of gradient in the [$\alpha$/Fe] ratio, either
positive or negative, and one as no gradient at all.

The build-up of the gradients is very fast and we predict
non significant evolution after the first 0.5 - 1 Gyr.

%\acknowledgements %%% Text of acknowledgements runs on after this command.

\end{document}